\documentclass[aps,twocolumn,superscriptaddress,footinbib,longbibliography]{revtex4-2}

\usepackage[utf8]{inputenc}
\usepackage{longtable}
\usepackage{placeins}
\usepackage{upgreek, textgreek}
\usepackage[dvips]{graphicx}
\usepackage{xcolor}
\usepackage{epsfig,graphicx,amsfonts,amsbsy}
\usepackage{amsmath,amsfonts,amsthm,amssymb}
\usepackage{appendix}
\usepackage{makeidx}
\usepackage{url}
\usepackage{verbatim}
\usepackage[bookmarksnumbered,pdfpagelabels=true,plainpages=false,colorlinks=true,linkcolor=blue,citecolor=red,urlcolor=blue]{hyperref}
\usepackage[rightcaption]{sidecap}
\usepackage{array}
\usepackage{multirow}
\usepackage{tabularx}
\usepackage{braket} 
\usepackage{dsfont}
\usepackage{bbold}
\usepackage{etoolbox}
\usepackage{diagbox}
\usepackage{float}
\usepackage{leftindex}

\usepackage{dcolumn}
\usepackage{bm}
\usepackage{blindtext}
\usepackage{comment}
\usepackage{physics}

\begin{document}

\title{Hidden quantum correlations in the ground states of quasiclassical spin systems}

\author{Levente  R{\'o}zsa}
\affiliation{Department of Theoretical Solid State Physics, HUN-REN Wigner Research Centre for Physics, H-1525 Budapest, Hungary}
\affiliation{Department of Theoretical Physics, Budapest University of Technology and Economics, H-1111 Budapest, Hungary}

\author{Dennis Wuhrer}
\affiliation{Department of Physics, University of Konstanz, 78457 Konstanz, Germany}

\author{Sebasti{\'a}n A. D{\'i}az}
\affiliation{Department of Physics, University of Konstanz, 78457 Konstanz, Germany}

\author{Ulrich Nowak}
\affiliation{Department of Physics, University of Konstanz, 78457 Konstanz, Germany}

\author{Wolfgang Belzig}
\affiliation{Department of Physics, University of Konstanz, 78457 Konstanz, Germany}

\date{\today}

\begin{abstract}

Frustrated spin models may lead to the formation of both classical non-collinear spin structures and unique quantum phases 
including highly entangled quantum spin liquids. Here, we study the entanglement and spatial quantum correlations in linear spin-wave theory around a classical spin-spiral ground state. We find that the entanglement between pairs of sites is short-ranged, and is completely absent in certain cases. In contrast, the entanglement hidden in multi-site clusters is peaked close to phase transitions and shows an asymptotic behavior modulated by the period of the magnetic structure. These findings motivate further exploring the connection in the entanglement properties of fully quantum and of quasiclassical spin models.

\end{abstract}

\maketitle

\section{Introduction}

Strongly correlated quantum systems often give rise to exotic phases. The study of these phases has long been a strong driving force behind the development of analytical and numerical methods for treating correlated systems~\cite{White1992,White1993,Ors2019}, as well as for the design of experiments and quantum simulations for their observation~\cite{Georgescu2014}. These phases are typically characterized by strong quantum entanglement, which has been demonstrated to show a scaling behavior at quantum phase transitions~\cite{Osterloh2002}. Due to their entangled ground state, strongly correlated quantum systems have been put forward as a robust quantum computing platform~\cite{Brennen2008,Chen2009,Wei2011}. 

Spin models strongly coupled by exchange interactions provide a fertile ground for studying quantum correlations. Finding the 
ground state of such models is a daunting task, even for systems having a simple classical analogue such as the two-sublattice antiferromagnetic Heisenberg model. Exact solutions to this problem are available in one dimension for spin $S=1/2$ for certain parameter ranges~\cite{Bethe1931,Lieb1961,Majumdar1969}. Although the ground state of an antiferromagnetic chain can rarely be constructed for higher spins, it has been proven that its excitation gap vanishes for infinite chains for half-integer spin~\cite{Affleck1986}, while for integer spin the system is conjectured to remain gapped~\cite{Haldane1983a,Haldane1983b}. 
Numerical methods such as density-matrix renormalization group~\cite{White1992,White1993} are especially well suited for one-dimensional systems. 
Antiferromagnetic interactions also give rise to strongly correlated magnetic phases in higher dimensions, for example quantum spin liquids characterized by a high degree of entanglement~\cite{Balents2010,Savary2016,Zhu2015}. However, two- and three-dimensional systems are difficult to treat numerically in the fully quantum case, especially 
their correlations over long distances that require large system sizes.

Magnons or spin waves describe the excitations of magnetically ordered systems. In contrast to the strongly correlated phases, they may be calculated by an expansion starting from a non-fluctuating classical spin configuration, obtained in the $S\rightarrow\infty$ limit of the quantum model. Nevertheless, magnons follow bosonic statistics, and as such may be utilized in quantum information and communication based on continuous variables~\cite{Braunstein2005,Eisert2003,Adesso2007}. 
The entanglement between magnons in single-mode ferromagnets or two-mode antiferromagnets and their correlations with cavity photons or phonons has been studied recently~\cite{Kamra2019,Zhang2019,Li2019,Yuan2020,Yuan2021,Yuan2022}. However, the limited number of modes considered in these previous works restricted the investigations to macroscopic quantum effects.

Non-collinear spin configurations such as spin spirals or skyrmions are present in a wide range of materials, and they possess intriguing topological and transport properties~\cite{Bergmann2014,Back2020,GoebelPR2021}. 
The spatial correlations in these textures are more complex than in ferromagnets or collinear antiferromagnets which can be described by a handful of modes. Non-collinear spin structures are stabilized by a competition between different interaction terms which often 
result in the formation of strongly correlated quantum phases as well~\cite{Balents2010}. The typical length scale of these structures ranges from a few nanometers to micrometers, but the investigation of their quantum properties~\cite{Psaroudaki2021,Haller2024} has been hampered by the fact that available numerical methods only enable the study of the smallest sizes in this range even for two-dimensional systems~\cite{Lohani2019,Sotnikov2021,Siegl2022,Haller2022,Joshi2024}. While the properties of magnons in these systems has been widely studied in the classical limit~\cite{Garst2017}, their quantum properties have been investigated much less, with the exception of magnon squeezing~\cite{Wuhrer2023}.

Here, we determine the entanglement between sites in non-collinear magnetic ground states based on quasiclassical spin-wave theory, illustrated on the example of spin spiral states. Going beyond macroscopic quantum effects, we consider all magnon modes in the system to study how the non-collinear structure modulates the entanglement on the atomic level. 
We demonstrate that the entanglement between pairs of sites is short-ranged and completely vanishes in certain parameter regimes, but find enhanced entanglement hidden in multi-site clusters close to phase transitions and at long distances. 
Finally, we discuss the analogy between the $S\rightarrow\infty$ quasiclassical and the $S=1/2$ ultraquantum limit.

\section{Methods}

Following Ref.~\cite{Wuhrer2023}, we consider the spin Hamiltonian
\begin{align}
\hat{\mathcal{H}}=\frac{1}{2}\sum_{i,j}J_{ij}\hat{\boldsymbol{S}}_{i}\hat{\boldsymbol{S}}_{j}+\frac{1}{2}\sum_{i,j}\boldsymbol{D}_{ij}\left(\hat{\boldsymbol{S}}_{i}\times\hat{\boldsymbol{S}}_{j}\right)-\mu_{\textrm{B}}gB_{\boldsymbol{n}}\sum_{i}\boldsymbol{n}\hat{\boldsymbol{S}}_{i},\label{eq:HamJD}
\end{align}
where $\hat{\boldsymbol{S}}_{i}$ is the spin operator at lattice site $i$, the $J_{ij}$ are Heisenberg exchange interactions, $\boldsymbol{D}_{ij}$ is the Dzyaloshinsky--Moriya vector~\cite{DMI_Dzyaloshinskii,DMI_Moriya} between sites $i$ and $j$, $B_{\boldsymbol{n}}$ is the projection of the external magnetic field on the direction $\boldsymbol{n}$, $\mu_{\textrm{B}}$ is the Bohr magneton and $g$ is the gyromagnetic factor. 

First, we determine the classical ground state of the Hamiltonian in Eq.~\eqref{eq:HamJD}. This corresponds to finding the unit vectors $\boldsymbol{S}_{i}^{(0)}$ satisfying
\begin{align}
\boldsymbol{S}_{i}^{(0)}\times\frac{\partial \mathcal{H}}{\partial \boldsymbol{S}_{i}^{(0)}}=0\label{eq:torque}
\end{align}
at each lattice site, where $\mathcal{H}$ is a classical Hamiltonian obtained after replacing the spin operators by classical vectors. The condition \eqref{eq:torque} expresses that the torque acting on each spin vanishes in the ground state. The same Hamiltonian may have many different local energy minima all satisfying Eq.~\eqref{eq:torque}, and the spin-wave expansion may be performed around any of these local minima. However, in the following we assume that $\boldsymbol{S}_{i}^{(0)}$ is also the global minimum of the Hamiltonian, i.e., the ground state. In the general case, the directions $\boldsymbol{S}_{i}^{(0)}$ differ between the sites, corresponding to a non-collinear ground state, and the ground state can only be found numerically, for example via the solution of the Landau--Lifshitz--Gilbert equation~\cite{Landau35,Gilbert2004}. In Eq.~\eqref{eq:HamJD}, two terms may prefer non-collinear states. First, the competition between ferromagnetic ($J_{ij}<0$) and antiferromagnetic ($J_{ij}>0$) Heisenberg interactions with different neighbors gives rise to frustration and a modulation of the spin structure. Second, the Dzyaloshinsky--Moriya interaction prefers a perpendicular alignment of the spins, thus it also competes with the Heisenberg interactions. The ground states formed by these two mechanisms may have the same spin structure, but their symmetry properties differ, which has a strong influence on their correlation measures as will be discussed below.

We identify the classical ground state with the quantum state $\left|0\right>_{\textrm{spin}}=\prod_{i}\left|S_{i}\right>$, a product of maximally polarized eigenstates along the direction of $\boldsymbol{S}_{i}^{(0)}$. Being a product state, $\left|0\right>_{\textrm{spin}}$ is completely uncorrelated between the different lattice sites. However, typically $\left|0\right>_{\textrm{spin}}$ does not correspond to the ground state or even any eigenstate of the quantum Hamiltonian. An approximate quantum ground state may be obtained by determining the low-energy excitations around this classical ground state within linear spin-wave theory. 
Therefore, we introduce a right-handed orthonormal basis $\left\{\boldsymbol{e}_{i,1},\boldsymbol{e}_{i,2},\boldsymbol{e}_{i,3}=\boldsymbol{S}_{i}^{(0)}\right\}$ at each lattice site, and express the components of the spin operators in this basis $\hat{S}_{i,\alpha}=\hat{\boldsymbol{S}_{i}}\cdot\boldsymbol{e}_{i,\alpha}$ using the linearized Holstein--Primakoff transformation~\cite{Holstein_Primakoff_Trafo},
\begin{align}
\hat{S}_{i,1}=&\sqrt{\frac{S}{2}}\left(\hat{a}_{i}+\hat{a}^{\dag}_{i}\right)=\sqrt{S}\hat{q}_{i},\label{eq:HPtrans1}\\
\hat{S}_{i,2}=&-\textrm{i}\sqrt{\frac{S}{2}}\left(\hat{a}_{i}-\hat{a}^{\dag}_{i}\right)=\sqrt{S}\hat{p}_{i},\label{eq:HPtrans2}\\
\hat{S}_{i,3}=&S-\hat{a}^{\dag}_{i}\hat{a}_{i},\label{eq:HPtrans3}
\end{align}
where $S$ is the spin quantum number, $\left[\hat{a}_{i},\hat{a}^{\dag}_{j}\right]=\delta_{ij}$ are bosonic creation and annihilation operators, and $\left[\hat{q}_{i},\hat{p}_{j}\right]=\textrm{i}\delta_{ij}$ are conjugate generalized coordinates and momenta. Substituting this transformation into the Hamiltonian in Eq.~\eqref{eq:HamJD} and keeping terms only up to second order in the bosonic operators yields
\begin{align}
\hat{\mathcal{H}}_{\textrm{SW}}=E_{0}+\frac{1}{2}\sum_{i,j=1}^{N}\left(\omega_{ij}\hat{a}^{\dag}_{i}\hat{a}_{j}+\mu_{ij}\hat{a}_{i}\hat{a}_{j}+\textrm{h. c.}\right),\label{eq:HSW}
\end{align}
where $N$ is the number of sites, $E_{0}$ is the classical ground-state energy, and $\omega_{ij}$ and $\mu_{ij}$ are coefficients of the spin-wave Hamiltonian.

In this approximation, the classical ground state in spin space $\left|0\right>_{\textrm{spin}}$ is replaced by the bosonic vacuum $\left|0\right>_{\textrm{osc}}=\prod_{i}\left|0_{i}\right>$ with $\hat{a}_{i}\left|0_{i}\right>=0$, i.e., a product of the ground states of independent harmonic oscillators. This state remains uncorrelated between sites, and is typically still not an eigenstate of $\hat{\mathcal{H}}_{\textrm{SW}}$ because the pairing terms with coefficients $\mu_{ij}$ do not conserve the boson number. To find the ground state $\left|\textrm{GS}\right>$ of $\hat{\mathcal{H}}_{\textrm{SW}}$, we look for bosonic operators $\hat{\boldsymbol{\alpha}}=\left(\hat{\alpha}_{1},\dots,\hat{\alpha}_{N},\hat{\alpha}^{\dag}_{1},\dots,\hat{\alpha}^{\dag}_{N}\right)$ as linear combinations of the original operators $\hat{\boldsymbol{a}}=\left(\hat{a}_{1},\dots,\hat{a}_{N},\hat{a}^{\dag}_{1},\dots,\hat{a}^{\dag}_{N}\right)$,
\begin{align}
\hat{\boldsymbol{\alpha}}=\mathcal{U}^{-1}\hat{\boldsymbol{a}},\label{eq:aalphatrans}
\end{align}
with the condition that they diagonalize the spin-wave Hamiltonian,
\begin{align}
\hat{\mathcal{H}}_{\textrm{SW}}=E_{\textrm{GS}}+\sum_{k}\Omega_{k}\hat{\alpha}^{\dag}_{k}\hat{\alpha}_{k}.\label{eq:HSWeig}
\end{align}
This requires finding the eigenvalues and eigenvectors of a matrix containing the $\omega_{ij}$ and $\mu_{ij}$ coefficients. 
The squeezed ground state satisfies $\hat{\alpha}_{k}\left|\textrm{GS}\right>=0$, and is consequently an eigenstate of $\hat{\mathcal{H}}_{\textrm{SW}}$ with energy $E_{\textrm{GS}}<E_{0}=\leftindex_{\textrm{osc}}{\left<0\right|}\hat{\mathcal{H}}_{\textrm{SW}}\left|0\right>_{\textrm{osc}}$. Note that finding the fully quantum ground state of Eq.~\eqref{eq:HamJD} in a space of dimension $\left(2S+1\right)^{N}$, where $N$ is the number of lattice sites, is numerically a very challenging problem. In the linear spin-wave approximation, the problem simplifies to finding the eigenvalues and eigenvectors of a matrix of size $2N\times2N$, enabling the treatment of larger systems, particularly in higher dimensions. 

The correlations between the sites are contained in the covariance matrix of the coordinates and momenta,
\begin{align}
\gamma_{IJ}=\frac{1}{2}\left<\textrm{GS}\left|\left\{\hat{\zeta}_{I},\hat{\zeta}^{\dag}_{J}\right\}\right|\textrm{GS}\right>,\label{eq:covmat}
\end{align}
where $\hat{\zeta}_{I}=\hat{q}_{i}$ for $1\le I=i\le N$ and $\hat{\zeta}_{I}=\hat{p}_{i}$ for $N+1\le I=i+N\le 2N$. The covariance matrix completely characterizes the ground state, because the ground state of coupled harmonic oscillators is always a so-called Gaussian state~\cite{Eisert2003} where the expectation values of single bosonic operators vanish. In the $\hat{\alpha}_{I}$ operators diagonalizing the spin-wave Hamiltonian with $\hat{\alpha}_{I}=\hat{\alpha}_{i}$ for $1\le I=i\le N$ and $\hat{\alpha}_{I}=\hat{\alpha}^{\dag}_{i}$ for $N+1\le I=i+N\le 2N$, the covariance matrix may be simply expressed as
\begin{align}
\gamma^{\alpha}_{IJ}&=\left<\textrm{GS}\left|\hat{\alpha}_{I}\hat{\alpha}^{\dag}_{J}\right|\textrm{GS}\right>-\frac{1}{2}\left<\textrm{GS}\left|\left[\hat{\alpha}_{I},\hat{\alpha}^{\dag}_{J}\right]\right|\textrm{GS}\right>\nonumber\\&=\delta_{IJ}\mathbb{I}\left(I\le N\right)-\frac{1}{2}\left(-1\right)^{\mathbb{I}\left(I> N\right)}\delta_{IJ}\nonumber\\&=\delta_{IJ}\mathbb{I}\left(I\le N\right)-\frac{1}{2}\Sigma_{IJ},\label{eq:covmatalpha}
\end{align}
where $\mathbb{I}$ is the indicator function. This may be transformed to $\gamma^{a}$, the covariance matrix of the $\hat{a}_{I}$ bosonic operators, using the transformation matrix
\begin{align}
\mathcal{U}^{-1}=\left[\begin{array}{cc}V & W \\ W^{*} & V^{*}\end{array}\right]\label{eq:U}
\end{align}
from Eq.~\eqref{eq:aalphatrans}. Since the transformation must preserve the bosonic commutation relations, i.e., $\mathcal{U}\Sigma\mathcal{U}^{\dag}=\Sigma$ with $\Sigma$ from Eq.~\eqref{eq:covmatalpha}, $\mathcal{U}^{-1}$ is a symplectic matrix which can be expressed using two $N\times N$ matrices $V$ and $W$ as given in Eq.~\eqref{eq:U}. These matrices further satisfy the conditions
\begin{align}
VV^{\dag}-WW^{\dag}=&I,\label{eq:VW1}\\
VW^{T}-WV^{T}=&0,\\
V^{\dag}V-\left(W^{\dag}W\right)^{*}=&I,\\
V^{\dag}W-W^{T}V^{*}=&0.\label{eq:VW2}
\end{align}
Applying the transformations Eq.~\eqref{eq:aalphatrans}, then Eqs.~\eqref{eq:HPtrans1} and \eqref{eq:HPtrans2} to Eq.~\eqref{eq:covmatalpha}, and using the conditions above, the covariance matrix in the canonical coordinates may be expressed as
\begin{align}
\gamma=\frac{1}{2}\left[\begin{array}{cc}U_{-}^{\dag}U_{-} & \textrm{i}\left(U_{-}^{\dag}U_{+}-I\right) \\ -\textrm{i}\left(U_{+}^{\dag}U_{-}-I\right) & U_{+}^{\dag}U_{+}\end{array}\right],\label{eq:covmatcalc}
\end{align}
with $U_{\pm}=V\pm W$.

The covariance matrix $\gamma$ describes the correlations between transversal spin components at certain lattice sites, and as such should be more readily accessible to local measurements than the eigenmodes $\alpha$ typically corresponding to extended states. If the variance $\gamma_{i,i}$ of the generalized coordinate $\hat{q}_{i}$ differs from the variance $\gamma_{i+N,i+N}$ of the generalized momentum $\hat{p}_{i}$, the ground state is squeezed. However, there is a U$\left(1\right)$ gauge freedom in the definition of the covariance matrix, since the $\boldsymbol{e}_{i,1}$ and $\boldsymbol{e}_{i,2}$ directions used to define $\hat{q}_{i}$ and $\hat{p}_{i}$ in Eqs.~\eqref{eq:HPtrans1} and \eqref{eq:HPtrans2} can be freely rotated around the classical equilibrium spin direction $\boldsymbol{S}_{i}^{(0)}=\boldsymbol{e}_{i,3}$ while keeping the vectors orthonormal. It is useful to introduce the gauge-invariant symplectic eigenvalues 
of the covariance matrix, by using the eigenvalues $\pm\textrm{i}c_{i}$ of the matrix $-\textrm{i}\sigma^{y}\gamma$, where the $\sigma^{y}=\left[\begin{array}{cc}0 &-\textrm{i} \\ \textrm{i} & 0\end{array}\right]$ Pauli matrix acts on the subspace of coordinates and momenta. The symplectic nature of the transformation $\mathcal{U}^{-1}$ ensures that the $c_{i}$ are all non-negative real numbers.

To characterize the correlations in the spin system, we select a subsystem in real space, separate it into two parts, and from the symplectic eigenvalues calculate gauge-invariant correlation measures between the parts. The von Neumann entropy of the state with covariance matrix $\gamma$ may be expressed as~\cite{Adesso2007}
\begin{align}
S_{V}\left(\gamma\right)=\sum_{i=1}^{N}f\left(2c_{i}\right),\label{eq:vNent}
\end{align}
with
\begin{align}
f\left(x\right)=\frac{x+1}{2}\ln\left(\frac{x+1}{2}\right)-\frac{x-1}{2}\ln\left(\frac{x-1}{2}\right).\label{eq:vNentfun}
\end{align}
The relative information in a bipartite system based on the von Neumann entropy is defined as
\begin{align}
S_{V,\textrm{rel}}=S_{V}\left(\gamma_{1}\right)+S_{V}\left(\gamma_{2}\right)-S_{V}\left(\gamma\right),\label{eq:vNrelinf}
\end{align}
where $\gamma_{1}$ and $\gamma_{2}$ are the covariance matrices of the two subsystems. These are simply submatrices of the covariance matrix of the complete system, $\gamma_{1/2,IJ}=\left\{\gamma_{IJ}|I,J\in \mathcal{I}_{1/2}\right\}$, where the index sets $\mathcal{I}_{1/2}$ contain the degrees of freedom included in the first or the second subsystem. These expressions are equivalent for Gaussian states to the common definitions of the von Neumann entropy and relative information based on the density matrix and its partial traces, but the latter are more complicated to calculate in the infinite-dimensional Hilbert spaces considered here. The von Neumann entropy remains subadditive, which means that the relative information is non-negative. For a pure state, all of the symplectic eigenvalues are $c_{i}=1/2$, and the von Neumann entropy vanishes. Therefore, the relative information may be used as a correlation measure, which for the two subsystems of a pure state measures purely quantum correlations or entanglement. The von Neumann entropy and the related R\'{e}nyi entropies calculated from spin-wave theory have been previously applied to study entanglement in collinear antiferromagnets~\cite{Song2011,Laflorencie2015} and spin spirals~\cite{Bauer2020}, for bipartitions of the whole system in the pure ground state. However, if $\gamma$ describes a mixed state of a subsystem, then $c_{i}\ge 1/2$ holds, and the relative information also contains classical correlations.

A more useful measure for separating quantum and classical correlations from each other in mixed states is the logarithmic negativity~\cite{Vidal2002}, which was also demonstrated to be an entanglement monotone~\cite{Plenio2005}. An important result in identifying separable or unentangled mixed states is the Peres-Horodecki criterion~\cite{Peres1996,Horodecki1996}. For a given bipartition into subsystems 1 and 2, the partial transpose $\varrho^{T_{1}}_{(i_{1}i_{2}),(j_{1}j_{2})}=\varrho_{(j_{1}i_{2}),(i_{1}j_{2})}$ of the density matrix being also a density matrix, namely a positive semidefinite matrix which preserves the trace of 1 of the original density matrix, is a necessary condition for separability. While no sufficient condition for separability exists in the general case, the positivity of the partial transpose of the density matrix has been proven to be a sufficient condition in certain cases. For Gaussian states, these include all bipartitions where one of the subsystems consists of a single harmonic oscillator~\cite{Simon2000}, satisfied by all pairs and clusters we will consider here. On the level of the covariance matrix, partial transposition corresponds to reversing the sign of the momenta $\hat{p}_{i}$ in one of the subsystems in Eq.~\eqref{eq:covmat}. Using the symplectic eigenvalues $\tilde{c}_{i}$ of the covariance matrix of the partial transposed system $\tilde{\gamma}$, the logarithmic negativity may be defined as
\begin{align}
E_{\mathcal{N}}=\sum_{i}\max\{0,-\log_{2}\left(2\tilde{c}_{i}\right)\}.\label{eq:logneg}
\end{align}
The logarithmic negativity will be positive if $\tilde{c}_{i}<1/2$ holds for any of the symplectic eigenvalues, which implies that the partial transpose of the density matrix is not positive (since $\tilde{c}_{i}\ge 1/2$ holds for density matrices), meaning that the state is not separable for the bipartitions mentioned above. In summary, when any subsystem is separated into a single harmonic oscillator and its environment, the logarithmic negativity will only be positive for entangled states, and its magnitude may be used as a measure of the quantum correlations in the system.

\section{Results}

\subsection{Application to conical spin spirals}

We will consider conical spin spiral ground states of Eq.~\eqref{eq:HamJD}, for which the covariance matrix $\gamma$ in Eq.~\eqref{eq:covmat} may be calculated analytically. The calculations will be performed for a two-dimensional triangular lattice. Following Ref.~\cite{Wuhrer2023}, we will find the ground state by first setting the magnetic field to zero, and considering the harmonic spin spiral $\boldsymbol{S}_{i}^{(0)}=\cos\left(\boldsymbol{q}_{0}\boldsymbol{R}_{i}\right)\boldsymbol{e}_{1}+\sin\left(\boldsymbol{q}_{0}\boldsymbol{R}_{i}\right)\boldsymbol{e}_{2}$, where $\boldsymbol{R}_{i}$ is the position of site $i$, and $\boldsymbol{e}_{1}$ and $\boldsymbol{e}_{2}$ are orthonormal vectors. The wave vector $\boldsymbol{q}_{0}$ and the normal vector of the spiral $\boldsymbol{n}=\boldsymbol{e}_{1}\times\boldsymbol{e}_{2}$ are chosen such that they minimize $\tilde{J}_{\boldsymbol{q}}$ calculated from the Fourier transform of the interactions,
\begin{align}
J_{\boldsymbol{q}}=&\sum_{\boldsymbol{R}_{i}-\boldsymbol{R}_{j}}\textrm{e}^{-\textrm{i}\boldsymbol{q}\cdot\left(\boldsymbol{R}_{i}-\boldsymbol{R}_{j}\right)}J_{ij},\\
D_{\boldsymbol{q}}^{\boldsymbol{n}}=&\sum_{\boldsymbol{R}_{i}-\boldsymbol{R}_{j}}\textrm{e}^{-\textrm{i}\boldsymbol{q}\cdot\left(\boldsymbol{R}_{i}-\boldsymbol{R}_{j}\right)}\boldsymbol{D}_{ij}\cdot\boldsymbol{n},\\
\tilde{J}_{\boldsymbol{q}}=&J_{\boldsymbol{q}}+\textrm{i}D_{\boldsymbol{q}}^{\boldsymbol{n}}.
\end{align}
Note that the rotational plane with normal vector $\boldsymbol{n}$ is defined by the Dzyaloshinsky--Moriya interactions, but is independent of the Heisenberg interactions. After $\boldsymbol{q}_{0}$ and $\boldsymbol{n}$ are fixed, the external field is oriented along $\boldsymbol{n}$, which leads to a finite magnetization by forcing the spin directions onto the surface of a cone, but does not distort the harmonic structure of the spiral.

We fix the gauge in the calculations by selecting the orthonormal basis used in Eqs.~\eqref{eq:HPtrans1}-\eqref{eq:HPtrans3} as
\begin{align}
\boldsymbol{e}_{i,1}=&\left[\begin{array}{c}\sin\left(\boldsymbol{q}_{0}\boldsymbol{R}_{i}\right)\cos\vartheta\\\cos\left(\boldsymbol{q}_{0}\boldsymbol{R}_{i}\right)\cos\vartheta\\\sin\vartheta\end{array}\right],\\
\boldsymbol{e}_{i,2}=&\left[\begin{array}{c}-\cos\left(\boldsymbol{q}_{0}\boldsymbol{R}_{i}\right)\\\sin\left(\boldsymbol{q}_{0}\boldsymbol{R}_{i}\right)\\0\end{array}\right],\\
\boldsymbol{e}_{i,3}=&\left[\begin{array}{c}\sin\left(\boldsymbol{q}_{0}\boldsymbol{R}_{i}\right)\sin\vartheta\\\cos\left(\boldsymbol{q}_{0}\boldsymbol{R}_{i}\right)\sin\vartheta\\\cos\vartheta\end{array}\right],
\end{align}
with the components expressed in the global right-handed basis $\left\{\boldsymbol{e}_{1},\boldsymbol{e}_{2},\boldsymbol{n}\right\}$, and $\vartheta$ denoting the opening angle of the cone determined by the strength of the external field, $\cos\vartheta=\mu_{\textrm{B}}gB_{\boldsymbol{n}}/\left(S\tilde{J}_{\boldsymbol{0}}-S\tilde{J}_{\boldsymbol{q}}\right)$~\cite{Wuhrer2023}.

After performing the Holstein--Primakoff transformation in Eqs.~\eqref{eq:HPtrans1}-\eqref{eq:HPtrans3}, the spin-wave Hamiltonian may be written in a simple form following Fourier transformation of the bosonic operators,
\begin{align}
\hat{a}_{\boldsymbol{q}}=&\frac{1}{\sqrt{N}}\sum_{i}\textrm{e}^{-\textrm{i}\boldsymbol{q}\cdot\boldsymbol{R}_{i}}\hat{a}_{i},\\
\hat{a}^{\dag}_{\boldsymbol{q}}=&\frac{1}{\sqrt{N}}\sum_{i}\textrm{e}^{\textrm{i}\boldsymbol{q}\cdot\boldsymbol{R}_{i}}\hat{a}^{\dag}_{i}.
\end{align}
Introducing the coefficients
\begin{align}
D_{0}\left(\boldsymbol{q}\right)=&S\left(1-\frac{\sin^{2}\vartheta}{2}\right)\frac{1}{2}\left(\tilde{J}_{\boldsymbol{q}_{0}+\boldsymbol{q}}+\tilde{J}_{\boldsymbol{q}_{0}-\boldsymbol{q}}\right)\nonumber\\&+S\frac{\sin^{2}\vartheta}{2}\frac{1}{2}\left(\tilde{J}_{\boldsymbol{q}}+\tilde{J}_{-\boldsymbol{q}}\right)-S\tilde{J}_{\boldsymbol{q}_{0}},\label{eq:D0}\\
D_{\textrm{nr}}\left(\boldsymbol{q}\right)=&S\cos\vartheta\frac{1}{2}\left(\tilde{J}_{\boldsymbol{q}_{0}+\boldsymbol{q}}-\tilde{J}_{\boldsymbol{q}_{0}-\boldsymbol{q}}\right),\\
D_{\textrm{a}}\left(\boldsymbol{q}\right)=&-S\frac{\sin^{2}\vartheta}{2}\frac{1}{2}\left(\tilde{J}_{\boldsymbol{q}_{0}+\boldsymbol{q}}+\tilde{J}_{\boldsymbol{q}_{0}-\boldsymbol{q}}\right)+S\frac{\sin^{2}\vartheta}{2}\frac{1}{2}\left(\tilde{J}_{\boldsymbol{q}}+\tilde{J}_{-\boldsymbol{q}}\right),\label{eq:Da}
\end{align}
the spin-wave Hamiltonian in Eq.~\eqref{eq:HSW} may be rewritten in a block-diagonal form of $2\times 2$ blocks,
\begin{align}
\hat{\mathcal{H}}_{\textrm{SW}}=&\tilde{E}_{0}+\frac{1}{2}\sum_{\boldsymbol{q}}\left[\left(D_{0}\left(\boldsymbol{q}\right)+D_{\textrm{nr}}\left(\boldsymbol{q}\right)\right)\hat{a}^{\dag}_{\boldsymbol{q}}\hat{a}_{\boldsymbol{q}}+D_{\textrm{a}}\left(\boldsymbol{q}\right)\hat{a}^{\dag}_{\boldsymbol{q}}\hat{a}^{\dag}_{-\boldsymbol{q}}\right.\nonumber\\&\left.+D_{\textrm{a}}\left(\boldsymbol{q}\right)\hat{a}_{-\boldsymbol{q}}\hat{a}_{\boldsymbol{q}}+\left(D_{0}\left(\boldsymbol{q}\right)-D_{\textrm{nr}}\left(\boldsymbol{q}\right)\right)\hat{a}_{-\boldsymbol{q}}\hat{a}^{\dag}_{-\boldsymbol{q}}\right].
\end{align}
Note that all coefficients are real in this gauge, and the relations $D_{0}\left(\boldsymbol{q}\right)=D_{0}\left(-\boldsymbol{q}\right)$, $D_{\textrm{nr}}\left(\boldsymbol{q}\right)=-D_{\textrm{nr}}\left(-\boldsymbol{q}\right)$, and $D_{\textrm{a}}\left(\boldsymbol{q}\right)=D_{\textrm{a}}\left(-\boldsymbol{q}\right)$ hold. The influence of the magnetic field is included in the $\vartheta$ parameter related to the opening angle of the spiral cone. 
Thus, the operators belonging to the eigenmodes of the system may be written as
\begin{align}
\hat{\alpha}_{\boldsymbol{q}}=v_{\boldsymbol{q}}\hat{a}_{\boldsymbol{q}}+w_{\boldsymbol{q}}\hat{a}^{\dag}_{-\boldsymbol{q}},
\end{align}
with the coefficients
\begin{align}
v_{\boldsymbol{q}}=\sqrt{\frac{1}{2}\left(\frac{D_{0}\left(\boldsymbol{q}\right)}{\sqrt{D^{2}_{0}\left(\boldsymbol{q}\right)-D^{2}_{\textrm{a}}\left(\boldsymbol{q}\right)}}+1\right)},\\
w_{\boldsymbol{q}}=\sqrt{\frac{1}{2}\left(\frac{D_{0}\left(\boldsymbol{q}\right)}{\sqrt{D^{2}_{0}\left(\boldsymbol{q}\right)-D^{2}_{\textrm{a}}\left(\boldsymbol{q}\right)}}-1\right)}.
\end{align}

Note that the matrices introduced in Eq.~\eqref{eq:U} are diagonal in Fourier space, $V_{\boldsymbol{q}\boldsymbol{q}'}=v_{\boldsymbol{q}}\delta_{\boldsymbol{q}\boldsymbol{q}'}$ and $W_{\boldsymbol{q}\boldsymbol{q}'}=w_{\boldsymbol{q}}\delta_{\boldsymbol{q}\boldsymbol{q}'}$, the elements are symmetric under switching the sign of $\boldsymbol{q}$, and are chosen to be real valued. They are normalized according to the relations in Eq.~\eqref{eq:VW1}-\eqref{eq:VW2}. Performing inverse Fourier transformation and using Eq.~\eqref{eq:covmatcalc}, the covariance matrix may be written in the compact form of
\begin{align}
&\boldsymbol{\gamma}_{ij}=\left[\begin{array}{cc}\gamma_{i,j}&\gamma_{i,j+N}\\\gamma_{i+N,j}&\gamma_{i+N,j+N}\end{array}\right]\nonumber\\=&\sum_{\boldsymbol{q}}\left[\begin{array}{cc}\sqrt{\frac{D_{0}\left(\boldsymbol{q}\right)-D_{\textrm{a}}\left(\boldsymbol{q}\right)}{D_{0}\left(\boldsymbol{q}\right)+D_{\textrm{a}}\left(\boldsymbol{q}\right)}}&0\\0&\sqrt{\frac{D_{0}\left(\boldsymbol{q}\right)+D_{\textrm{a}}\left(\boldsymbol{q}\right)}{D_{0}\left(\boldsymbol{q}\right)-D_{\textrm{a}}\left(\boldsymbol{q}\right)}}\end{array}\right]\frac{\cos\left[\boldsymbol{q}\cdot\left(\boldsymbol{R}_{i}-\boldsymbol{R}_{j}\right)\right]}{2N}.\label{eq:covmatspiral}
\end{align}

The covariance matrix only depends on the relative position of the two sites, which is a consequence of Fourier transformation being suitable for diagonalizing the spin-wave Hamiltonian. Furthermore, the covariance matrix is independent of the spin quantum number $S$, as can be seen after expressing the coefficients in the matrix using Eqs.~\eqref{eq:D0} and \eqref{eq:Da},
\begin{align}
D_{0}\left(\boldsymbol{q}\right)-D_{\textrm{a}}\left(\boldsymbol{q}\right)=&S\frac{1}{2}\left(\tilde{J}_{\boldsymbol{q}_{0}+\boldsymbol{q}}+\tilde{J}_{\boldsymbol{q}_{0}-\boldsymbol{q}}\right)-S\tilde{J}_{\boldsymbol{q}_{0}},\label{eq:D0mDa}\\
D_{0}\left(\boldsymbol{q}\right)+D_{\textrm{a}}\left(\boldsymbol{q}\right)=&S\left(1-\sin^{2}\vartheta\right)\frac{1}{2}\left(\tilde{J}_{\boldsymbol{q}_{0}+\boldsymbol{q}}+\tilde{J}_{\boldsymbol{q}_{0}-\boldsymbol{q}}\right)\nonumber\\&+S\sin^{2}\vartheta\frac{1}{2}\left(\tilde{J}_{\boldsymbol{q}}+\tilde{J}_{-\boldsymbol{q}}\right)-S\tilde{J}_{\boldsymbol{q}_{0}},\label{eq:D0pDa}
\end{align}
and taking their ratio. Since $\boldsymbol{q}_{0}$ was found by minimizing $\tilde{J}_{\boldsymbol{q}}$ in order to obtain the ground state, clearly both $D_{0}\left(\boldsymbol{q}\right)-D_{\textrm{a}}\left(\boldsymbol{q}\right)$ and $D_{0}\left(\boldsymbol{q}\right)+D_{\textrm{a}}\left(\boldsymbol{q}\right)$ are non-negative. It can also be seen from the expressions that $D_{0}\left(\boldsymbol{q}\right)-D_{\textrm{a}}\left(\boldsymbol{q}\right)$ vanishes for $\boldsymbol{q}=\boldsymbol{0}$. This is connected to a zero mode of the system, found by calculating the eigenfrequencies in Eq.~\eqref{eq:HSWeig},
\begin{align}
\Omega_{\boldsymbol{q}}=D_{\textrm{nr}}\left(\boldsymbol{q}\right)+\sqrt{D^{2}_{0}\left(\boldsymbol{q}\right)-D^{2}_{\textrm{a}}\left(\boldsymbol{q}\right)}.
\end{align}
The eigenfrequency also vanishes for $\boldsymbol{q}=\boldsymbol{0}$ because of $D_{\textrm{nr}}\left(\boldsymbol{0}\right)=0$. This mode is connected to a U$\left(1\right)$ symmetry broken by the choice of the ground state: 
a continuous translation of the spiral along the wave vector, or a change in the phase of the spiral, 
results in a state with the same energy. 
This implies that the summation over the $\boldsymbol{q}$ points in Eq.~\eqref{eq:covmatspiral} must be performed carefully to avoid the singularity at $\boldsymbol{q}=\boldsymbol{0}$. For this purpose, the summation is performed numerically over a regular grid in $\boldsymbol{q}$ space which does not include the $\boldsymbol{q}=\boldsymbol{0}$ point, similarly to the Monkhorst--Pack method~\cite{Monkhorst1976,Pack1977} but adapted to preserve a $C_{3\textrm{v}}$ symmetry of the grid points around $\boldsymbol{q}=\boldsymbol{0}$ on the triangular lattice. Expanding the Fourier transforms of the interaction parameters yields $D_{0}\left(\boldsymbol{q}\right)-D_{\textrm{a}}\left(\boldsymbol{q}\right)\propto\boldsymbol{q}^{2}$ apart from specific values of the interaction coefficients. 
This means that the singularity $\sqrt{D_{0}\left(\boldsymbol{q}\right)+D_{\textrm{a}}\left(\boldsymbol{q}\right)}/\sqrt{D_{0}\left(\boldsymbol{q}\right)-D_{\textrm{a}}\left(\boldsymbol{q}\right)}\propto q^{-1}$ is integrable in at least two dimensions, so the calculations should converge as the mesh is refined.

Due to the simple structure of the covariance matrix, the symplectic eigenvalues for one and two lattice sites may also be expressed in a compact form,
\begin{align}
c_{i}=\sqrt{\gamma_{i,i}\gamma_{i+N,i+N}},\label{eq:ci}
\end{align}
and
\begin{align}
c^{(1)}_{ij}=\sqrt{\left(\gamma_{i,i}+\gamma_{i,j}\right)\left(\gamma_{i+N,i+N}+\gamma_{i+N,j+N}\right)},\label{eq:cij1}\\
c^{(2)}_{ij}=\sqrt{\left(\gamma_{i,i}-\gamma_{i,j}\right)\left(\gamma_{i+N,i+N}-\gamma_{i+N,j+N}\right)},\label{eq:cij2}
\end{align}
with the coefficients defined in Eq.~\eqref{eq:covmatspiral}. These expressions may be used to calculate the two-site relative information based on the von Neumann entropy in Eq.~\eqref{eq:vNrelinf}. The partial transposition required for the logarithmic negativity in Eq.~\eqref{eq:logneg} corresponds to switching the sign of $\gamma_{i+N,j+N}$ in the two-site expression.

\subsection{Covariance matrix\label{sec:3B}}

We will compare the correlation measures between two types of spin spirals. The first type is stabilized by the next-nearest-neighbor antiferromagnetic Heisenberg interaction $J_{2}>0$, while the Dzyaloshinsky-Moriya interaction is set to zero. In the second type, the nearest-neighbor Dzyaloshinsky-Moriya interaction $D_{1}$ is taken into account, while $J_{2}$ is turned off. The nearest-neighbor Heisenberg interaction $J_{1}$ will be allowed to be either ferromagnetic or antiferromagnetic in both cases, and we will use the reduced parameters $\textrm{atan2}\left(J_{2},J_{1}\right)$ and $\textrm{atan2}\left(J_{1},D_{1}\right)$ to interpolate between these two limits.

\begin{figure}
\begin{centering}
\includegraphics[width=\columnwidth]{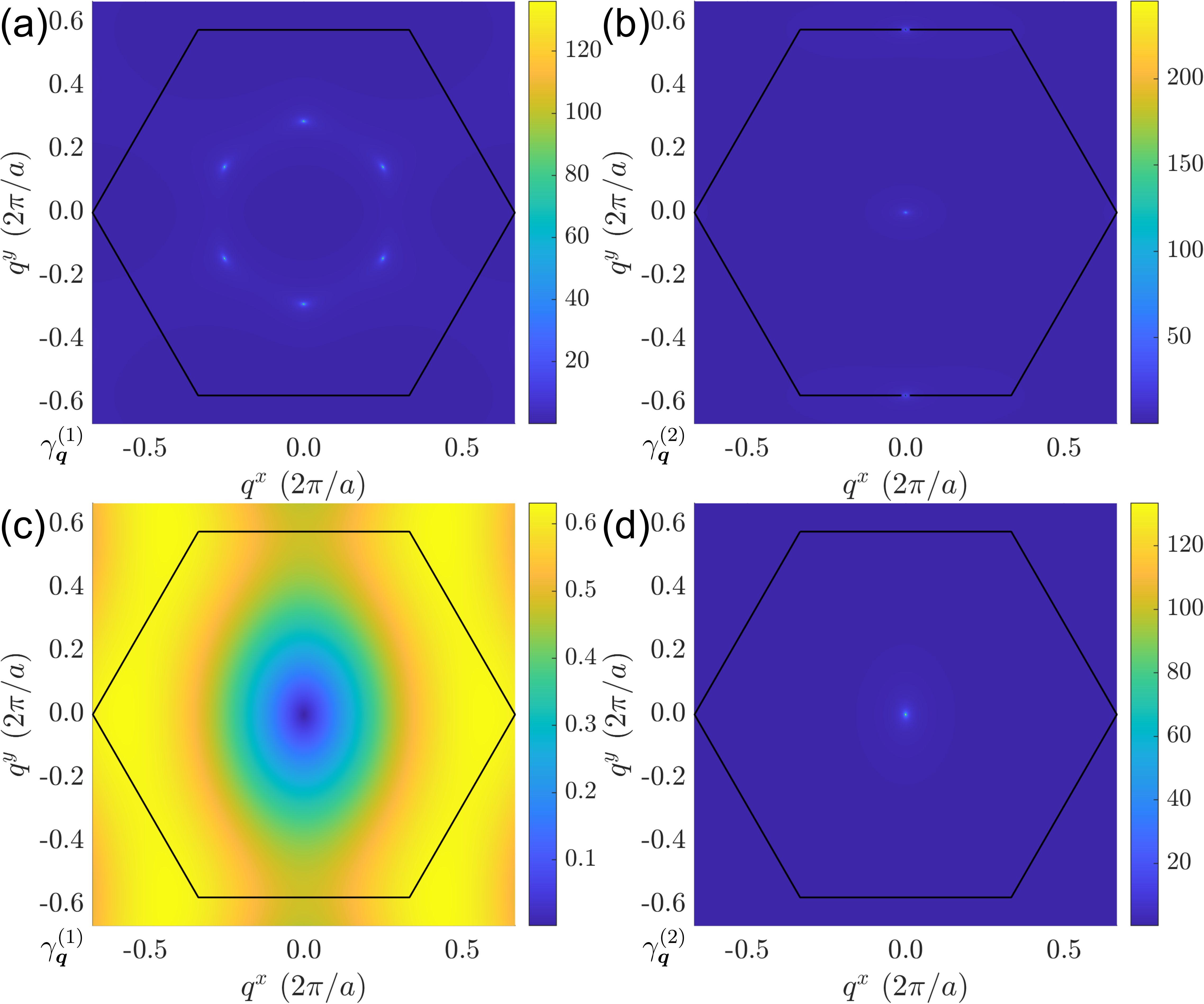}
\par\end{centering}
\caption{Elements of the covariance matrix in Fourier space. (a) $\gamma^{(1)}_{\boldsymbol{q}}$ in Eq.~\eqref{eq:gamma1q} and (b) $\gamma^{(2)}_{\boldsymbol{q}}$ in Eq.~\eqref{eq:gamma2q} calculated for the parameters $J_{2}=-J_{1}>0,D_{1}=0$. (c) $\gamma^{(1)}_{\boldsymbol{q}}$ and (d) $\gamma^{(2)}_{\boldsymbol{q}}$ calculated for the parameters $J_{2}=J_{1}=0,D_{1}>0$. The black hexagon illustrates the Brillouin zone, consisting of an $N=513\times 513$ grid. The distribution of $\gamma^{(1)}_{\boldsymbol{q}}$ is radically different because of the different symmetries of the two models.\label{fig:fig5}
}
\end{figure}

The elements of the covariance matrix in Fourier space,
\begin{align}
\gamma^{(1)}_{\boldsymbol{q}}=\frac{1}{2}\sqrt{\frac{D_{0}\left(\boldsymbol{q}\right)-D_{\textrm{a}}\left(\boldsymbol{q}\right)}{D_{0}\left(\boldsymbol{q}\right)+D_{\textrm{a}}\left(\boldsymbol{q}\right)}},\label{eq:gamma1q}\\
\gamma^{(2)}_{\boldsymbol{q}}=\frac{1}{2}\sqrt{\frac{D_{0}\left(\boldsymbol{q}\right)+D_{\textrm{a}}\left(\boldsymbol{q}\right)}{D_{0}\left(\boldsymbol{q}\right)-D_{\textrm{a}}\left(\boldsymbol{q}\right)}},\label{eq:gamma2q}
\end{align}
are illustrated in Fig.~\ref{fig:fig5} for these two scenarios. 
The translational zero mode at $\boldsymbol{q}=\boldsymbol{0}$ shows up as a singularity in $\gamma^{(2)}_{\boldsymbol{q}}$ both in Fig.~\ref{fig:fig5}(b) and (d). 
If only isotropic exchange interactions are taken into account, and the spiral is planar ($\sin\vartheta=1$), then $D_{0}\left(\boldsymbol{q}\right)+D_{\textrm{a}}\left(\boldsymbol{q}\right)$ in Eq.~\eqref{eq:D0pDa} will vanish for $\boldsymbol{q}=\boldsymbol{q}_{0}$ and all symmetrically equivalent points satisfying $\tilde{J}_{\boldsymbol{q}_{\alpha}}=\tilde{J}_{\boldsymbol{q}_{0}}$. This can be explained by the O(3) rotational symmetry of the Heisenberg Hamiltonian broken by the choice of the ground state, where any global spin rotation results in an energetically equivalent state. On the level of the spin-wave expansion around a selected ground state, this manifests itself as an additional zero mode, corresponding to changing the rotational plane of the spiral. 
This zero mode leads to singularities in $\gamma^{(1)}_{\boldsymbol{q}}$, as shown in Fig.~\ref{fig:fig5}(a). However, the rotational plane of the spiral becomes fixed 
if an external magnetic field is applied or the Dzyaloshinsky--Moriya interaction is taken into account. 
This results in no peaks and a much narrower range of values in the colorbar in Fig.~\ref{fig:fig5}(c); note that $\gamma^{(1)}\left(\boldsymbol{q}\right)\equiv 1/2$ would be obtained in the uncorrelated case. 

For comparison, we mention that the 
two-mode squeezing parameters $r_{\boldsymbol{q}}$ connecting the eigenmodes with $\boldsymbol{q}$ and $-\boldsymbol{q}$ derived in Ref.~\cite{Wuhrer2023} may be expressed using the elements of the covariance matrix in Eqs.~\eqref{eq:gamma1q} and ~\eqref{eq:gamma2q} as
\begin{align}
\textrm{e}^{2r_{\boldsymbol{q}}}=\sqrt{\frac{D_{0}\left(\boldsymbol{q}\right)+\left|D_{\textrm{a}}\left(\boldsymbol{q}\right)\right|}{D_{0}\left(\boldsymbol{q}\right)-\left|D_{\textrm{a}}\left(\boldsymbol{q}\right)\right|}}=2\max\left\{\gamma^{(1)}_{\boldsymbol{q}},\gamma^{(2)}_{\boldsymbol{q}}\right\}.\label{eq:squeezing}
\end{align}

\subsection{Pairwise and multipartite entanglement}

\begin{figure}
\begin{centering}
\includegraphics[width=\columnwidth]{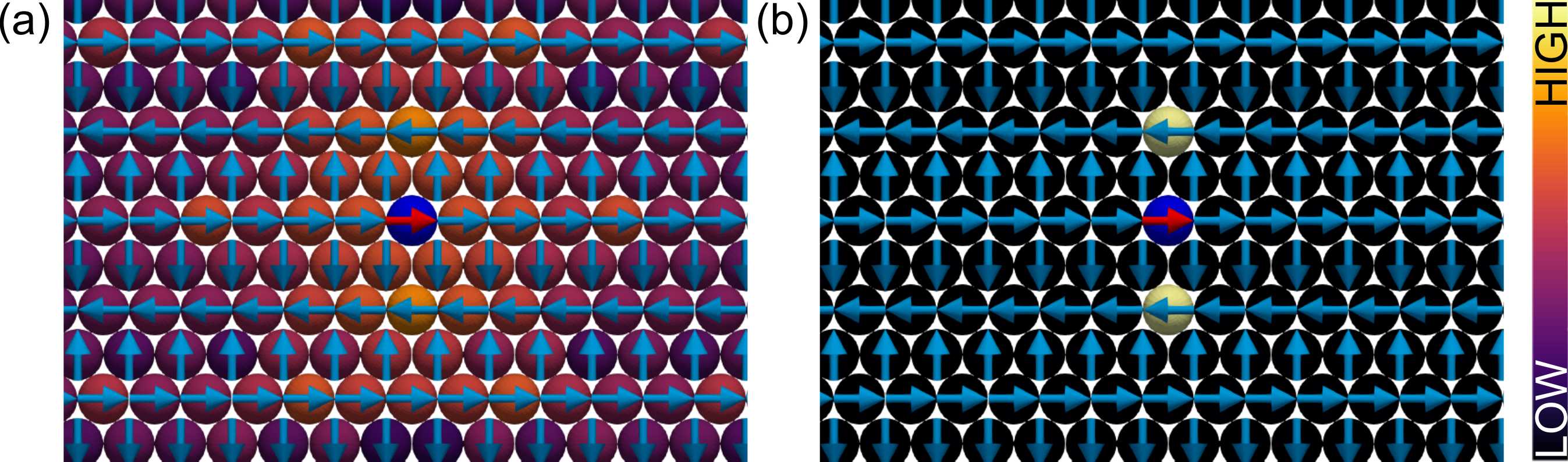}
\par\end{centering}
\caption{Spatial distribution of magnonic pair correlation measures in spin spirals. (a) Pairwise relative von Neumann information $S_{V,\textrm{rel,pw}}$ and (b) pairwise logarithmic negativity $E_{\mathcal{N},\textrm{pw}}$ between the center site (red arrow in blue sphere) and the other sites. The color of the spheres shows the magnitude of the correlations with that site, with the color scale being logarithmic. 
Arrows illustrate the classical spin spiral ground state. The interactions take the value $J_{2}=-J_{1}>0, D_{1}=0$, resulting in a spin spiral with 4 atoms period along the $y$ direction, $q_{0}^{y}=2\pi/\left(2\sqrt{3}a\right)$. The Brillouin zone integration was performed on an $N=513\times 513$ grid.
In stark contrast to the classical correlations in panel (a), the quantum correlations in panel (b) are anisotropic and short-ranged.
\label{fig:fig1}}
\end{figure}

From the covariance matrix, we calculated the correlation measures between pairs of sites. For the isotropic Heisenberg model with $J_{2}=-J_{1}>0,D_{1}=0$, these are illustrated in Fig.~\ref{fig:fig1} along with the calculated ground state. The relative information calculated from the von Neumann entropy $S_{V,\textrm{rel,pw}}$ for pairs of sites in Fig.~\ref{fig:fig1}(a) decays as the distance between the sites increases, but remains finite up to arbitrarily large distances. It can also be observed that the relative information is spatially modulated by the spin spiral structure, resulting in an oscillatory decay not only along the spiral modulation direction $y$, but along the perpendicular direction $x$ as well. In contrast, the logarithmic negativity calculated for pairs of sites is only finite for next-nearest-neighbors along the $y$ direction, while it is equal to zero for all other neighbors. 
The surprisingly short range of quantum correlations compared to classical correlations has also been observed for one-dimensional quantum spin chains in Ref.~\cite{Osterloh2002}, indicating a qualitative similarity between the entanglement calculated in the ground state of the spin Hamiltonian there 
and in the spin-wave Hamiltonian considered here.

\begin{figure}
\begin{centering}
\includegraphics[width=\columnwidth]{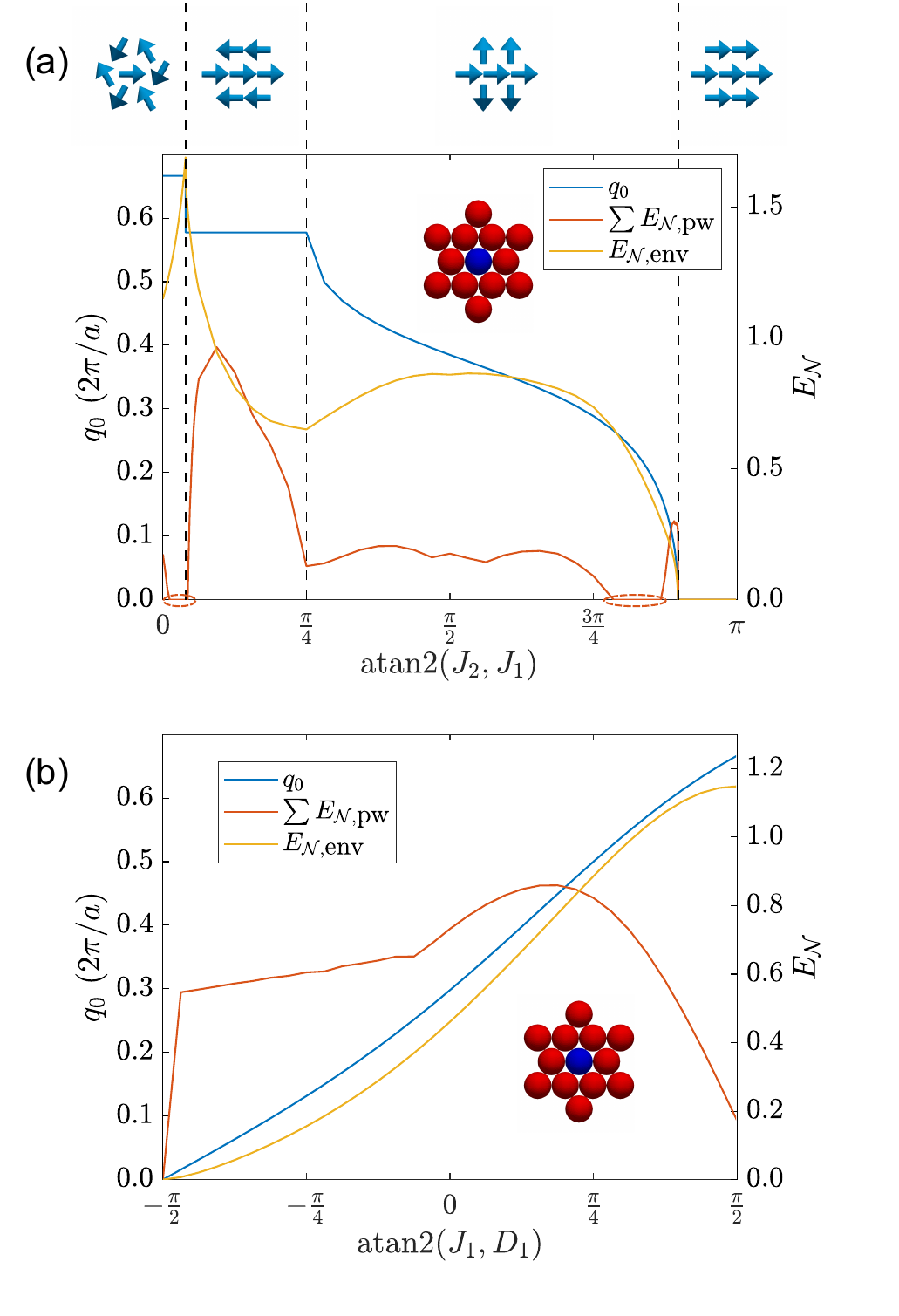}
\par\end{centering}
\caption{Phase diagram and logarithmic negativity. (a) Ordering wave vector of the spiral $q_{0}$, pairwise logarithmic negativity $\sum E_{\mathcal{N},\textrm{pw}}$ summed up over all sites, and logarithmic negativity $E_{\mathcal{N},\textrm{env}}$ with the environment as a function of the ratio of nearest-neighbor and next-nearest-neighbor exchange couplings, for $D_{1}=0$. Sketches at the top illustrate from left to right the N\'{e}el antiferromagnetic state, 
the row-wise antiferromagnetic state, 
the spin-spiral state, 
and the ferromagnetic state. 
The center site (blue) and the environment containing all nearest- and next-nearest-neighbour sites (red) used as a bipartition for calculating $E_{\mathcal{N},\textrm{env}}$ is also sketched. Dashed orange ellipses highlight the regions where $\sum E_{\mathcal{N},\textrm{pw}}$ vanishes but $E_{\mathcal{N},\textrm{env}}$ remains finite. (b) The same quantities as in panel (a) for the spin spirals stabilized by the Dzyaloshinsky--Moriya interaction, for $J_{2}=0$. The Brillouin zone integration was performed on an $N=513\times 513$ grid.
Multiple-site clusters, rather than pairs of sites, are required to reveal the hidden quantum entanglement.
\label{fig:fig2}}
\end{figure}

The entanglement is analyzed in an extended parameter space in Fig.~\ref{fig:fig2} for both types of spin spirals. In Fig.~\ref{fig:fig2}(a) we set $J_{2}>0,D_{1}=0$ and varied $J_{1}$ from positive to negative values. As illustrated by the blue curve, four different phases can be observed in the system, the boundaries between which may be determined by minimizing 
$\tilde{J}_{\boldsymbol{q}}$ 
with respect to the wave vector. For $J_{1}>0$ ($\textrm{atan2}\left(J_{2},J_{1}\right)<\pi/2$), the phases have been determined based on spin-wave theory in Ref.~\cite{Jolicouer1990}. For only nearest-neighbor antiferromagnetic interactions ($\textrm{atan2}\left(J_{2},J_{1}\right)=0$), the ground state is the three-sublattice Néel state with $q_{0}^{x}=4\pi/\left(3a\right),q_{0}^{y}=0$. This transforms into a row-wise antiferromagnetic state at $J_{2}/J_{1}= 1/8$ with $q_{0}^{x}=0,q_{0}^{y}=2\pi/\left(\sqrt{3}a\right)$. At the transition between the two types of antiferromagnetic states, these two states are degenerate with all other spin spirals with modulation vectors $\boldsymbol{q}_{0}$ along the boundary of the atomic Brillouin zone. The extensive ground-state degeneracy along a curve in the two-dimensional Brillouin zone gives rise to a spiral spin liquid at this point in the classical limit~\cite{Shang2022,Yan2024}. 
For $J_{2}/J_{1}\ge 1$ ($\textrm{atan2}\left(J_{2},J_{1}\right)\ge\pi/4$), the wave vector of the spiral starts to decrease while keeping its orientation along the next-nearest-neighbor direction in real space. The period of the spiral diverges at $J_{2}/J_{1}= -1/3$, and the system becomes ferromagnetically aligned. In Fig.~\ref{fig:fig2}(b), we set $J_{2}=0$, and considered the influence of the Dzyaloshinsky--Moriya interaction. In this case, the system transforms from the ferromagnetic state at $J_{1}<0,D_{1}=0$ to the Néel antiferromagnetic state $J_{1}>0,D_{1}=0$ through a spin spiral with ordering vector parallel to the $x$ axis as the angle $\textrm{atan2}\left(D_{1},J_{1}\right)$ is varied between $-\pi/2$ and $\pi/2$. The wave vector depends almost linearly on the angle, particularly close to the ferromagnetic state.

Although the spin spiral phase connects the ferromagnetic and the antiferromagnetic states in both models, the structure of the correlations is remarkably different. Calculating the logarithmic negativity between pairs of sites, and summing it up over all relative positions by taking advantage of its finite range, results in the orange curves in Fig.~\ref{fig:fig2}(a) and (b). The pairwise logarithmic negativity vanishes in the isotropic ferromagnetic state, since this is completely uncorrelated both in the spin-wave description and in the original spin Hamiltonian. In the $J_{1}-J_{2}$ model in Fig.~\ref{fig:fig2}(a), the entanglement between pairs 
is finite in the spin-spiral phase next to the phase transition into the ferromagnetic state, before 
vanishing in the range $2.47\le\textrm{atan2}\left(J_{2},J_{1}\right)\le2.72$. In this regime, the range of the pairwise logarithmic negativity is $0$, and no pairwise quantum correlations can be observed. For longer spiral periods just outside this region, the pairwise logarithmic negativity is only finite between nearest-neighbor sites along the modulation direction, while for shorter periods it is only finite for next-nearest neighbors, as illustrated in Fig.~\ref{fig:fig1}(b). This indicates that the value of the pairwise logarithmic negativity is sensitive to the period of the modulation. The logarithmic negativity again increases for shorter spin-spiral periods, and assumes even higher values in the row-wise antiferromagnetic phase. In contrast, the sum of the pairwise logarithmic negativity is finite everywhere in the $J_{1}-D_{1}$ model in Fig.~\ref{fig:fig2}(b), apart from the abrupt drop to zero in the ferromagnetic state; in particular, the logarithmic negativity between nearest-neighbor pairs also remains finite. In Fig.~\ref{fig:fig2}(a), this quantity again vanishes in the range $0.037\le\textrm{atan2}\left(J_{2},J_{1}\right)\le0.136$, including the classical spiral spin liquid point. To the left of this regime in the N\'{e}el state, the logarithmic negativity is finite for all nearest-neighbor pairs, while in the row-wise antiferromagnetic state immediately to the right of this region it is only finite between nearest neighbors with opposite spin directions. 
In Fig.~\ref{fig:fig2}(b), a decrease in the sum of the pairwise logarithmic negativity values is observed at higher wave vectors, but it does not vanish.


Since the spin-wave ground state remains correlated everywhere apart from the ferromagnetic state, we performed further calculations to uncover the correlations in the regimes where the sum of the pairwise logarithmic negativity values vanishes. For this purpose, we considered a 13-atom cluster sketched in Fig.~\ref{fig:fig2}, and calculated the logarithmic negativity over the bipartition into the central atom and all other atoms, as shown by the yellow curves. The logarithmic negativity calculated with this environment only vanishes in the ferromagnetic state, but it becomes finite in the regimes inside the correlated spin spiral and antiferromagnetic phases where no pairwise logarithmic negativity was found in Fig.~\ref{fig:fig2}(a). This means that the entanglement in these parameter ranges cannot be extracted when looking at only two lattice sites, instead it is hidden in clusters consisting of multiple sites. Well-known examples of analogous multipartite entanglement for three spin-$1/2$ particles include the GHZ state~\cite{Greenberger1989} and the W state~\cite{Duer2000}. Particularly interesting is the phase transition between the Néel and the row-wise antiferromagnetic states, where the pairwise logarithmic negativity vanishes but the value calculated for the larger cluster reaches a maximum when approaching from either side. The full Heisenberg spin Hamiltonian has long been considered as a highly entangled quantum-spin-liquid candidate in the vicinity of this classical phase transition~\cite{Zhu2015}, while in quasiclassical spin-wave theory 
we find an enhanced but exclusively multipartite entanglement. In Fig.~\ref{fig:fig2}(b), the logarithmic negativity calculated with the 12-site environment closely follows the monotonic change in $q_{0}$ with $\textrm{atan2}\left(J_{1},D_{1}\right)$, i.e., the degree of entanglement appears to be simply proportional to the angle between neighboring spins.

The difference between the two models concerning the distance dependence of the pairwise logarithmic negativity can be qualitatively understood based on the covariance matrices in Fourier space shown in Fig.~\ref{fig:fig5}. Due to the singularity of $\gamma^{(2)}_{\boldsymbol{q}}$ at $\boldsymbol{q}=\boldsymbol{0}$ caused by the translational mode, its Fourier transform at $\boldsymbol{R}_{i}-\boldsymbol{R}_{j}=\boldsymbol{0}$ $\gamma_{i+N,i+N}$ is typically considerably larger than $1/2$. The other on-site coefficient $\gamma_{i,i}$ is also quite large if there are singularities in $\gamma^{(1)}_{\boldsymbol{q}}$ as in the $J_{1}-J_{2}$ model in Fig.~\ref{fig:fig5}(a), but it is close to $1/2$ if $\gamma^{(1)}_{\boldsymbol{q}}$ is smooth as in the $J_{1}-D_{1}$ model in Fig.~\ref{fig:fig5}(c). Note that $\gamma_{i,i}$ would take the value $1/2$ in the uncorrelated case where $\gamma^{(1)}_{\boldsymbol{q}}\equiv 1/2$. Since the correlation matrix elements decrease rapidly with the distance between $i$ and $j$, at large distances one obtains $\tilde{c}^{(1)}_{ij}\approx \tilde{c}^{(2)}_{ij}\approx c_{i}>1/2$ from Eqs.~\eqref{eq:ci}-\eqref{eq:cij2}, and the logarithmic negativity between the two sites in Eq.~\eqref{eq:logneg} vanishes. Because both $\gamma_{i,i}$ and $\gamma_{i+N,i+N}$ are large in the isotropic model with nearest- and next-nearest-neighbor Heisenberg interactions $J_{1}$ and $J_{2}$, the logarithmic negativity vanishes for all pairs for certain parameter ranges. However, this is not observed for any parameter set if only nearest-neighbor Heisenberg and Dzyaloshinsky--Moriya interactions $J_{1}$ and $D_{1}$ are considered, where $\gamma^{(1)}_{\boldsymbol{q}}$ is not singular. Fundamentally, this effect may be attributed to the Dzyaloshinsky--Moriya interaction breaking the rotational symmetry of the plane of the spiral. However, note that considering a very small value of the Dzyaloshinsky--Moriya interaction is sufficient for breaking the symmetry, but it does not restore the pairwise logarithmic negativity unless it is sufficiently strong to compete with the Heisenberg interactions.

\subsection{Distance dependence of the entanglement}

\begin{figure}
\begin{centering}
\includegraphics[width=\columnwidth]{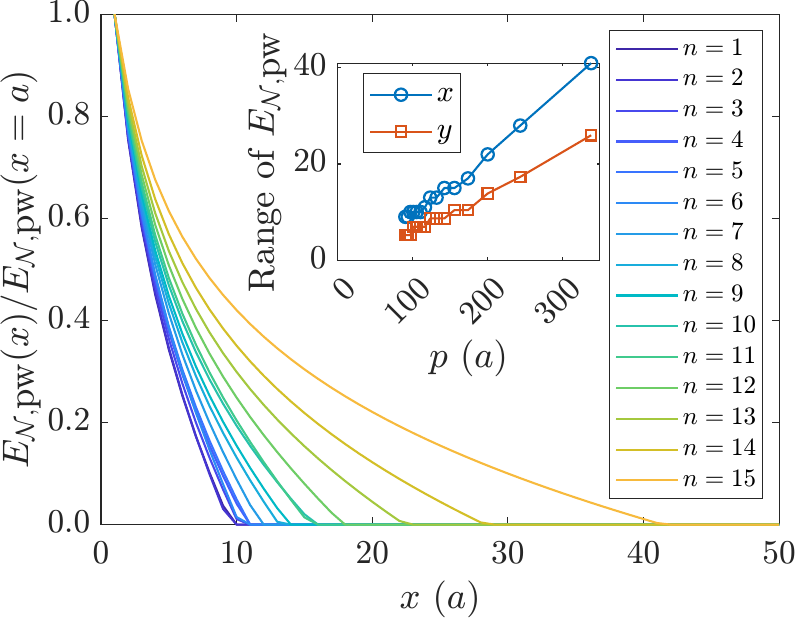}
\par\end{centering}
\caption{Pairwise logarithmic negativity close to the spin spiral-ferromagnetic transition. The pairwise logarithmic negativity $E_{\mathcal{N},\textrm{pw}}$ is calculated for neighbors along the $x$ direction, and normalized to the nearest-neighbor value. The interaction parameters are $\textrm{atan2}\left(J_{2},J_{1}\right)=919\pi/2^{10}+n\pi/2^{17},D_{1}=0$. Inset shows the range of $E_{\mathcal{N},\textrm{pw}}$ as a function of the spin-spiral period $p$. The Brillouin zone integration was performed on an $N=1026\times 1026$ grid.
The range of pairwise quantum correlations in spin spirals increases almost linearly with the spiral period.
\label{fig:fig3}}
\end{figure}

It was demonstrated above that the logarithmic negativity between pairs of sites is typically only finite between a few pairs, or it may completely vanish in certain parameter regimes. Here, we will investigate the asymptotic behavior of the quantum correlations in the model. First, the logarithmic negativity between pairs of sites 
close to the transition from the spin spiral to the ferromagnetic state is shown in Fig.~\ref{fig:fig3} for the $J_{1}-J_{2}$ model. The interactions took the values $\textrm{atan2}\left(J_{2},J_{1}\right)=919\pi/2^{10}+n\pi/2^{17}$ for different integer $n$ values as denoted in the legend, with higher $n$ values denoting spirals with longer periods closer to the phase transition. The range of this type of correlation increases with the period of the spiral approximately linearly, as shown in the inset. However, this range remains much shorter than the period. This type of quantum correlation extends up to larger distances perpendicular to the spiral modulation direction ($x$) than along the modulation direction ($y$). Note that the amplitude of the logarithmic negativity decreases as the phase transition is approached (see Fig.~\ref{fig:fig2}), but Fig.~\ref{fig:fig3} shows the curves normalized to the nearest-neighbor value for better visibility. A similar behavior is observed in the $J_{1}-D_{1}$ model close to the ferromagnetic limit at $\textrm{atan2}\left(J_{1},D_{1}\right)=-\pi/2$.

\begin{figure}
\begin{centering}
\includegraphics[width=0.8\columnwidth]{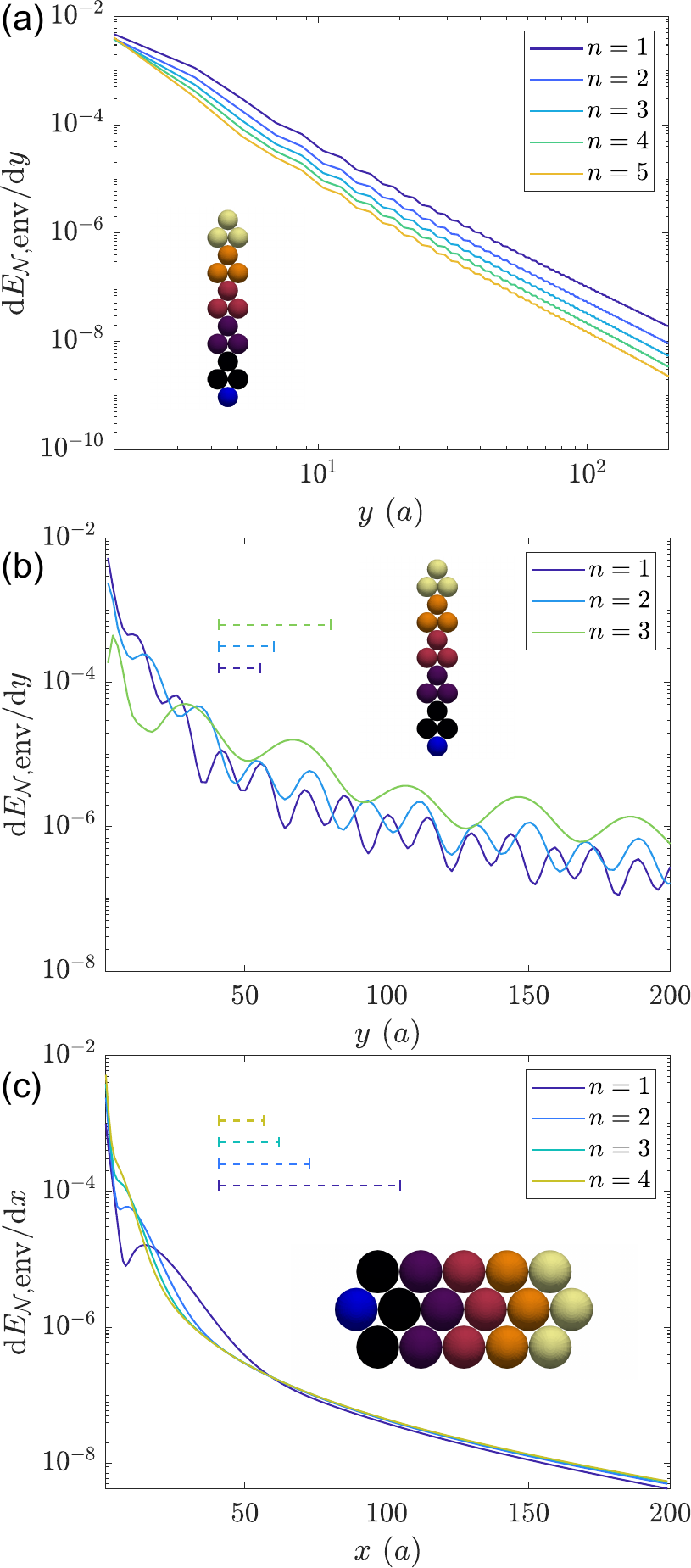}
\par\end{centering}
\caption{
Derivative of the logarithmic negativity with the environment along the modulation direction (a) close to the transition from the row-wise antiferromagnetic state to the N\'{e}el antiferromagnetic state ($\textrm{atan2}\left(J_{2},J_{1}\right)=5\pi/2^{7}+n\pi/2^{8},D_{1}=0$), and close to the transition from the spin-spiral state to the ferromagnetic state (b) for frustrated Heisenberg interactions ($\textrm{atan2}\left(J_{2},J_{1}\right)=1829\pi/2^{11}+n\pi/2^{9},D_{1}=0$) and (c) in the presence of the Dzyaloshinsky--Moriya interaction ($\textrm{atan2}\left(J_{1},D_{1}\right)=-\pi/2+n\pi/2^{5},J_{2}=0$). The logarithmic negativity was calculated for the bipartition on the lowest site (blue) and the environment (yellow to purple atoms). In each step, the environment was increased by three atoms denoted by the same color along the direction of the spiral modulation vector: along $y$ in panels (a) and (b) and along $x$ in panel (c). $\textrm{d}E_{\mathcal{N},\textrm{env}}/\textrm{d}y$ and $\textrm{d}E_{\mathcal{N},\textrm{env}}/\textrm{d}x$ are the logarithmic negativity values differentiated with respect to the size of the environment. 
Dashed lines in panels (b) and (c) illustrate the spin-spiral period $p$ for the given parameters. The Brillouin zone integration was performed on an $N=1026\times 1026$ grid.
The power-lay decay of $\textrm{d}E_{\mathcal{N},\textrm{env}}/\textrm{d}y$ in the vicinity of both transitions is consistent with a gapless excitation spectrum. 
The Goldstone mode associated with the free choice of the rotational plane of the spiral causes the spatial modulation in the $J_{1}-J_{2}$ model.
\label{fig:fig4}}
\end{figure}

The logarithmic negativity between pairs of sites is not suitable for studying the asymptotic behavior of quantum correlations, since its range is short compared to the characteristic length scale of the magnetic order in all parameter regimes. 
As an alternative measure, we considered the logarithmic negativity between a site and its increasingly larger environment along a certain direction, illustrated by the sketch in Fig.~\ref{fig:fig4}(a). The $E_{\mathcal{N},\textrm{env}}$ calculated this way monotonically increases with the size of the environment. We also calculated $\textrm{d}E_{\mathcal{N},\textrm{env}}/\textrm{d}y$, i.e., the increase in $E_{\mathcal{N},\textrm{env}}$ when three additional atoms are added to the cluster divided by $\Delta y=\sqrt{3}a$, and found that it converges to zero. Similarly to the decaying function $S_{V,\textrm{rel,pw}}$, $\textrm{d}E_{\mathcal{N},\textrm{env}}/\textrm{d}y$ may be considered as a correlation function, describing by how much the quantum correlations between a site and its environment increase by adding further sites to the cluster at a given distance. We considered a three-atom-wide environment because close to the transition between the two antiferromagnetic states, the logarithmic negativity does not only vanish for all pairs of sites but between sites and their one-atom-wide extensions as well. The parameters were varied according to the relation $\textrm{atan2}\left(J_{2},J_{1}\right)=5\pi/2^{7}+n\pi/2^{8}$ inside the row-wise antiferromagnetic state, where lower $n$ values are closer to the phase transition [cf. Fig.~\ref{fig:fig2}(a)]. 
It can be seen from the linear behavior on the log-log scale in Fig.~\ref{fig:fig4}(a) that $\textrm{d}E_{\mathcal{N},\textrm{env}}/\textrm{d}y$ may be well approximated by a power function. Systems with an ungapped excitation spectrum typically have correlation functions decaying according to a power law, while a gapped excitation spectrum usually results in an exponential decay. In the present system, the ungapped spectrum is caused by the translational symmetry 
of the spiral on the lattice, as discussed in Sec.~\ref{sec:3B}. 
The 
correlation functions decay slower closer to the transition point (i.e., smaller values of $n$); see Table~\ref{tab:fit1} 
for the 
fitted exponents of the power law. At the phase transition, 
the extensive ground-state degeneracy of the spiral spin liquid results in nodal lines in the magnon spectrum where the frequency vanishes, and a divergence of the correlations.

\begin{table}[!htbp]
\begin{minipage}{0.3\columnwidth}
\begin{ruledtabular}
\begin{tabular}{cr}
$n$ & $\eta$ \\
\hline
$1$ & $2.54\pm0.04$ \\
$2$ & $2.60\pm0.05$ \\
$3$ & $2.64\pm0.05$ \\
$4$ & $2.69\pm0.05$ \\
$5$ & $2.73\pm0.05$ 
\end{tabular}
\end{ruledtabular}
\end{minipage}
\caption{Exponents of the correlation function in the frustrated Heisenberg model in the row-wise antiferromagnetic state shown in Fig.~\ref{fig:fig4}(a). 
The fitted function was $\textrm{d}E_{\mathcal{N},\textrm{env}}/\textrm{d}y=ay^{-\eta}$. The interaction parameters are $\textrm{atan2}\left(J_{2},J_{1}\right)=5\pi/2^{7}+n\pi/2^{8},D_{1}=0$.}
\label{tab:fit1}
\end{table}

A similar power-law decay of $\textrm{d}E_{\mathcal{N},\textrm{env}}/\textrm{d}y$ is observed close to the transition from the spin spiral to the ferromagnetic state in the $J_{1}-J_{2}$ model for increasing $n$ values in $\textrm{atan2}\left(J_{2},J_{1}\right)=1829\pi/2^{11}+n\pi/2^{9}$ in Fig.~\ref{fig:fig4}(b). The spatial modulation of the correlation 
function is also apparent in this case, and the highest Fourier component of this modulation approximately corresponds to the period of the spin spiral along the considered $y$ direction, as illustrated by the dashed segments and the numerical fitting parameters of an oscillating power-law decay in Table~\ref{tab:fit2}. 
The modulation is not apparent 
in Fig.~\ref{fig:fig4}(a) because the period of the modulation in the row-wise antiferromagnetic state coincides with the next-nearest-neighbor distance along $y$, which is precisely the step size with which the environment is increased. However, these oscillations are also absent in Fig.~\ref{fig:fig4}(c) for the same quantity calculated in spin spirals stabilized by the Dzyaloshinsky--Moriya interaction, where the transition is approached as $n$ is decreased in $\textrm{atan2}\left(J_{1},D_{1}\right)=-\pi/2+n\pi/2^{5}$; see Table~\ref{tab:fit3} for the 
fitted exponents of the power law. 
Note that the Dzyaloshinsky-Moriya interaction changes the preferred modulation direction from $y$ to $x$, but the energy difference between the directions is small in this regime where the period is much longer than the lattice constant. 
The modulation can be understood based on the zero mode connected to the free choice of the rotational plane of the spiral in the $J_{1}-J_{2}$ model. Due to the singularity of $\gamma^{(1)}_{\boldsymbol{q}}$ at the points $\boldsymbol{q}_{\alpha}$ symmetry equivalent to $\boldsymbol{q}_{0}$ in Fig.~\ref{fig:fig5}(a), the covariance matrix elements in Eq.~\eqref{eq:covmatspiral} may be approximated as
\begin{align}
\gamma_{i,j}\approx\sum_{\alpha}f\left(\left|\boldsymbol{R}_{i}-\boldsymbol{R}_{j}\right|\right)\cos\left[\boldsymbol{q}_{\alpha}\cdot\left(\boldsymbol{R}_{i}-\boldsymbol{R}_{j}\right)\right]
\end{align}
by performing the summation around the singularities which give the largest contribution to the sum. Here, the $f\left(\left|\boldsymbol{R}_{i}-\boldsymbol{R}_{j}\right|\right)$ function shows a similar power-law decay with negligible oscillations around each singularity $\boldsymbol{q}_{\alpha}$, while the $\cos\left[\boldsymbol{q}_{\alpha}\cdot\left(\boldsymbol{R}_{i}-\boldsymbol{R}_{j}\right)\right]$ causes oscillations with the distance with the characteristic wave vectors of the spiral $\boldsymbol{q}_{\alpha}$. These modulations are also observable in other correlation measures, such as the relative von Neumann information calculated between pairs of sites shown in Fig.~\ref{fig:fig1}(a). However, the zero mode connected to the rotational plane of the spiral and the accompanying singularities in the correlations in Fig.~\ref{fig:fig5}(c) are absent 
in spin spirals stabilized by the Dzyaloshinsky--Moriya interaction, and the correlation function in Fig.~\ref{fig:fig4}(c) does not display periodic modulations. 

\begin{table}[!htbp]
\begin{minipage}{0.8\columnwidth}
\begin{ruledtabular}
\begin{tabular}{crrr}
$n$ & $\eta$ & $k$ $\left(2\pi/a\right)$ & $q_{0}$ $\left(2\pi/a\right)$\\
\hline
$1$ & $2.27\pm0.13$ & $0.0683\pm0.0004$ & $0.0682$\\
$2$ & $2.14\pm0.14$ & $0.0521\pm0.0005$ & $0.0517$\\
$3$ & $2.16\pm0.12$ & $0.0260\pm0.0004$ & $0.0254$
\end{tabular}
\end{ruledtabular}
\end{minipage}
\caption{Fitting parameters of the correlation function frustrated Heisenberg model in the spin-spiral state shown in Fig.~\ref{fig:fig4}(b). 
The fitted function was $\textrm{d}E_{\mathcal{N},\textrm{env}}/\textrm{d}y=\left[a+b\cos\left(ky+\varphi\right)\right]y^{-\eta}$. The interaction parameters are $\textrm{atan2}\left(J_{2},J_{1}\right)=1829\pi/2^{11}+n\pi/2^{9},D_{1}=0$. The wave number of the spiral $q_{0}$ is given for comparison with the modulation parameter $k$.}
\label{tab:fit2}
\end{table}

\begin{table}[!htbp]
\begin{minipage}{0.3\columnwidth}
\begin{ruledtabular}
\begin{tabular}{cr}
$n$ & $\eta$ \\
\hline
$1$ & $2.93\pm0.02$ \\
$2$ & $2.88\pm0.01$ \\
$3$ & $2.85\pm0.01$ \\
$4$ & $2.82\pm0.01$ 
\end{tabular}
\end{ruledtabular}
\end{minipage}
\caption{Exponents of the correlation functions in the Dzyaloshinsky--Moriya model in the spin-spiral state shown in Fig.~\ref{fig:fig4}(c). The fitted function was $\textrm{d}E_{\mathcal{N},\textrm{env}}/\textrm{d}x=ax^{-\eta}$. The interaction parameters are $\textrm{atan2}\left(J_{1},D_{1}\right)=-\pi/2+n\pi/2^{5},J_{2}=0$.}
\label{tab:fit3}
\end{table}

\section{Conclusion}

In summary, we investigated correlations in non-collinear magnetic ground states based on quasiclassical linear spin-wave theory. We used the relative information calculated from the von Neumann entropy to describe both classical and quantum correlations, and the logarithmic negativity for calculating exclusively quantum correlations or entanglement. In a nearest- and next-nearest-neighbor Heisenberg model on the triangular lattice, we found that the entanglement between pairs of sites has short range, and it vanishes completely around the 
spiral spin liquid point between the row-wise antiferromagnetic and the Néel antiferromagnetic phases, as well as in a regime inside the spin-spiral phase. In these regions, the entanglement is hidden in multisite clusters. For spin spirals stabilized by the Dzyaloshinsky--Moriya interaction, we did not observe such a complete absence of pairwise quantum correlations in any parameter range, and found that the entanglement is typically larger for higher angles between neighboring spins. The maximal distance at which pairwise entanglement can be observed increases close to the transition to the ferromagnetic state in both considered models, but it stays shorter than the modulation period. We calculated the asymptotic behavior of the entanglement by studying clusters of increasing size, and found that the derivative of the logarithmic negativity with respect to the cluster size follows a power-law decay, 
with the exponent 
decreasing in magnitude towards the spiral spin liquid point where magnon correlations diverge. The power-law decay is modulated by the period of the spiral in the frustrated Heisenberg model, but not in spirals stabilized by the Dzyaloshinsky--Moriya interaction. Although the spin structures formed by the frustrated Heisenberg interactions and by the Dzyaloshinsky--Moriya interaction may be identical, the choice of the ground state breaks the O$\left(3\right)$ rotational symmetry of the model in the former case, while it only breaks the U$\left(1\right)$ symmetry connected to the translation of the spiral in the latter case. 
This difference in symmetry results in the 
deviations in the correlation measures between the two models mentioned above.

The short-range nature of the entanglement observed here is also prevalent in fully quantum spin models; for example, it underlies the area laws of entanglement growth with system size~\cite{Eisert2010}. It has been demonstrated in Ref.~\cite{Osterloh2002} that the entanglement may show singular behavior close to quantum phase transitions; here, a similar effect is observed for the logarithmic negativity for larger environments 
close to the classical phase transition between the row-wise and Néel antiferromagnetic states. 
Established experimental methods and first-principles calculations may be utilized to design the magnetic interactions for engineering non-collinear structures~\cite{Back2020}, enabling the exploration of an extended parameter space in real materials. State-of-the-art imaging techniques already enable resolving the magnetic structure on the atomic scale~\cite{Wiesendanger2009,Schlenhoff2019,Tanigaki2024}, and might be developed further towards measuring spin fluctuations. It has been proposed recently that the magnon correlations in ferromagnetic or antiferromagnetic ground states may be detected by coupling them to qubits~\cite{Roemling2023,Roemling2024}, which may be generalized to local measurements in non-collinear spin structures. The low computational complexity of spin-wave theory should facilitate the exploration of entanglement properties 
at long distances in two- or three-dimensions, 
and the investigation of utilizing quasiclassical systems for quantum information processing. 

\begin{acknowledgments}

This work was financially supported by the Deutsche Forschungsgemeinschaft (DFG, German Research Foundation) via the Collaborative Research Center SFB 1432 (Project No.~425217212), by the National Research, Development, and Innovation Office (NRDI) of Hungary under Project Nos. K131938 and FK142601, by the Ministry of Culture and Innovation and the National Research, Development and Innovation Office within the Quantum Information National Laboratory of Hungary (Grant No. 2022-2.1.1-NL-2022-00004), and by the Hungarian Academy of Sciences via a J\'{a}nos Bolyai Research Grant (Grant No. BO/00178/23/11).

\end{acknowledgments}

%

\end{document}